\documentclass[10pt, twocolumn, aps, prd, nofootinbib, superscriptaddress, preprintnumbers]{revtex4-1}

\pdfoutput=1

\usepackage{amsmath, amssymb, amsfonts}
\usepackage{bm,bbm}
\usepackage{graphicx, graphics}
\usepackage{color}
\usepackage{hyperref}
\usepackage{cleveref}

\newcommand{\lb}[1]{\label{#1}}
\newcommand{\eq}[1]{(\ref{e.#1})}
\newcommand{\Eq}[1]{Eq.~(\ref{e.#1})}

\newcommand{\Ref}[1]{Ref.~\cite{#1}}
\newcommand{\Refs}[1]{Refs.~\cite{#1}}
\newcommand{\Fig}[1]{Fig.~\ref{f.#1}}

\newcommand{\App}[1]{Appendix~\ref{a.#1}}
\newcommand{\Sec}[1]{Section~\ref{s.#1}}

\newcommand{\beq}{\begin{equation}\begin{aligned}}
\newcommand{\eeq}{\end{aligned}\end{equation}}
\newcommand{\beqa}[1]{\begin{equation}\begin{alignedat}{#1}}
\newcommand{\eeqa}{\end{alignedat}\end{equation}}

\newcommand{\fr}[2]{\dfrac{#1}{#2}}
\newcommand{\wba}[1]{\overline{#1}}
\newcommand{\wtd}[1]{\widetilde{#1}}

\newcommand{\lt}{\left}
\newcommand{\rt}{\right}
\newcommand{\bra}{\langle}
\newcommand{\ket}{\rangle}

\newcommand{\lr}{\leftrightarrow}

\newcommand{\lsim}{\lesssim}

\newcommand{\dt}{\!\cdot\!}

\newcommand{\del}{\partial}
\newcommand{\dd}{\mathrm{d}}

\newcommand{\e}{\mathrm{e}}

\newcommand{\al}{\alpha}
\newcommand{\be}{\beta}
\newcommand{\ga}{\gamma}
\newcommand{\Ga}{\Gamma}
\newcommand{\de}{\delta}
\newcommand{\ep}{\epsilon}
\newcommand{\la}{\lambda}
\newcommand{\sg}{\sigma}

\newcommand{\co}{{\mathcal O}}

\newcommand{\SU}{\mathrm{SU}}

\newcommand{\PE}{{\phantom=}}
\newcommand{\ds}[1]{\displaystyle{#1}}

\newcommand{\nbar}{{\bar{n}}}
\newcommand{\nubar}{{\bar{\nu}}}
\newcommand{\Bbar}{{\bar{B}}}
\newcommand{\Lbar}{{\bar{L}}}
\newcommand{\hard}{H}
\newcommand{\soft}{Z_\text{\tiny S}}
\newcommand{\pT}{p_\mathrm{T}}
\newcommand{\als}{\alpha_\mathrm{s}}
\newcommand{\xibar}{{\bar{\xi}}}
\newcommand{\ptv}{p_\mathrm{T}^\text{veto}}
\newcommand{\albar}{{\bar{\alpha}}}
\newcommand{\ptvbar}{\bar{p}_{\mathrm T}^\text{veto}}
\newcommand{\Lp}{L_\perp}
\newcommand{\Lpbar}{\bar{L}_\perp}
\newcommand{\fbar}{{\bar{f}}}
\newcommand{\gbar}{{\bar{g}}}
\newcommand{\Ls}{L_\text{\tiny S}}
\newcommand{\LM}{L_M}
\newcommand{\nuinitial}{\nu_\text{\tiny B}}
\newcommand{\nubarinitial}{\nubar_\text{\tiny B}}
\newcommand{\nusinitial}{\nu_\text{\tiny S}}
\newcommand{\nusbarinitial}{\nubar_\text{\tiny S}}
\newcommand{\as}{a_\mathrm{s}}

\begin{document}

\title{Re-emergence of rapidity scale uncertainty in SCET}

\author{Prerit Jaiswal}
\affiliation{Department of Physics, Syracuse University, Syracuse, NY 13244, USA}

\author{Takemichi Okui}
\affiliation{Department of Physics, Florida State University, Tallahassee, FL 32306, USA}

\begin{abstract}
The artificial separation of a full-theory mode into distinct collinear and soft modes in SCET leads to divergent integrals over rapidity, which are not present in the full theory. Rapidity divergence introduces an additional scale into the problem, giving rise to its own renormalization group with respect to this new scale. Two contradicting claims exist in the literature concerning rapidity scale uncertainty. One camp has shown that the results of perturbative calculations depend on the precise choice of rapidity scale. The other has derived an all-order factorization formula with no dependence on rapidity scale, by using a form of analytic regulator to regulate rapidity divergences. We deliver a simple resolution to this controversy by deriving an alternative form of the all-order factorization formula with an analytic regulator that, despite being formally rapidity scale independent, reveals how rapidity scale dependence arises when it is truncated at a finite order in perturbation theory. With our results, one can continue to take advantage of the technical ease and simplicity of the analytic regulator approach while correctly taking into account rapidity scale dependence. As an application, we update our earlier study of $WW$ production with jet-veto by including rapidity scale uncertainty. While the central values of the predictions are unchanged, the scale uncertainties are increased and consistency between the NLL and NNLL calculations are improved.
\end{abstract}

\maketitle


\section{Introduction}
\lb{s.intro}
The paradigm and framework of effective field theory (EFT) is essential in our conceptual understanding of quantum field theory as well as in practical calculations.
In an EFT, by definition, we have removed (i.e., integrated out) 
the modes of the full theory that are necessarily highly off-shell 
with the given initial and final states,
so that the Lagrangian only contains the modes that can be on-shell 
and may produce singularities in loop and/or phase-space integrals. 
Also fundamental in EFT is to have small parameters besides coupling constants and assign to each field in the theory a definite power-counting rules in terms of those parameters,
so that we have manifestly well-controlled expansions at the Lagrangian level.
Often, defining such power-counting rules requires classifying the EFT modes into smaller groups of modes and introduce a separate field for each group.

Soft collinear effective theory 
(SCET)~\cite{Bauer:2000ew, Bauer:2000yr, Bauer:2001ct, Bauer:2001yt, Bauer:2002nz}
\cite{Beneke:2002ph, Beneke:2002ni}
is an excellent example of EFT with such further mode separations. In a typical application of SCET to collider physics, 
we have \emph{collinear modes,} 
i.e., the energetic modes nearly parallel to one beam direction, and 
\emph{anti-collinear modes,} i.e., those nearly parallel to the other beam.
It is also often convenient or necessary to introduce \emph{soft modes,} which are between collinear and anti-collinear modes.%
\footnote{We restrict our discussion to SCET$_{\text{II}}$-type observables, for which collinear, anti-collinear and soft modes suffice. 
Other observables may require different modes such as ultra-soft modes.}
To introduce our notation and nomenclature, 
let us quickly review the power-counting rules for those modes. 
Let $n$ and $\nbar$ be the 4-vectors $(1,0,0,1)$ and $(1,0,0,-1)$, respectively.
Any 4-vector $v$ can then be decomposed uniquely as $v^\mu = v_+ \, n^\mu /2 + v_- \,\nbar^\mu / 2 + v_\perp^\mu$, where $v_+ \equiv \ds{\nbar \dt v}$ and $v_- \equiv \ds{n \dt v}$, and $\ds{n \dt v_\perp} = \ds{\nbar \dt v_\perp} = 0$. 
Then, a collinear momentum $p$, an anti-collinear momentum $\bar{p}$, and a soft momentum $p_\mathrm{s}$ are characterized by the following scaling behavior in powers of a small power-counting parameter $\la \ll 1$:
\beq
p &= (p_+, p_-, p_\perp) \sim (1, \la^2, \la)M
\,,\\
\bar{p} &= (\bar{p}_+, \bar{p}_-, \bar{p}_\perp) \sim (\la^2, 1, \la)M
\,,\\
p_\mathrm{s} &= (p_{\mathrm{s}+}, p_{\mathrm{s}-}, p_{\mathrm{s}\perp}) \sim (\la, \la, \la)M
\,,\lb{e.power-counting}
\eeq
where $M$ is the characteristic high energy scale of the process in question.
Perturbative computation in SCET is a dual expansion in $\la$ and the coupling constant $\als$.
All the three modes have the same level of virtuality, $p^2 \sim \bar{p}^2 \sim p_\mathrm{s}^2 \sim \la^2 M^2$, 
but their rapidities $\eta$ (along the beam axis) are vastly different, i.e., $\e^\eta \sim \la^{-1}$, $\sim \la$, and $\sim 1$, respectively.
As noted earlier, 
the EFT should describe those three groups of modes by three separate fields $\phi(x)$, $\bar{\phi}(x)$, and $\phi_\mathrm{s}(x)$ (even if they all originally belong to the same field $\Phi(x)$ in the full theory) 
so that we can assign a definite power of $\la$ to each field, and thereby each term, in the Lagrangian.
   
Such separation of modes, however, can lead to additional divergences in loop and/or phase-space integrals~\cite{Collins:1981uk, Collins:1992tv, Collins:2008ht}. 
Just like we get ultraviolet (UV) divergences by ignoring the boundary (or cutoff) between the highly off-shell modes we have integrated out and the nearly on-shell modes we have kept, 
we get \emph{rapidity divergences} by neglecting the boundaries between collinear, soft, and anti-collinear modes.
Then, just like UV divergence leads to a renormalization group (RG) with respect to a scale parameter $\mu$, 
rapidity divergence leads to its own RG with respect to a different scale parameter $\nu$~\cite{Chiu:2007dg, Chiu:2011qc, Chiu:2012ir}.
As usual, the $\mu$ RG equations (RGEs) can be used to resum large logarithms arising from a (Lorentz invariant) hierarchy of scales in the problem.
Similarly,  
the $\nu$-RGEs can be used to resum large logarithms from a (frame-dependent) hierarchy of rapidity scales (as in \Eq{power-counting}).

A natural question, then, is whether there are additional scale uncertainties associated with the precise choice of initial and final values of $\nu$ in solving the $\nu$-RGEs, 
analogous to the familiar scale uncertainties from the choice of $\mu$.  
At first sight, it might appear that the $\nu$-RG should not constitute an \emph{independent} source of uncertainties.
Firstly, both $\mu$- and $\nu$-RGEs resum the same large logarithms of the form $\log(1/\la)$, because the theory actually contains only one large ratio of scales, $1/\la$,
as can be seen in the scaling laws~\eq{power-counting}.
Since there is only one kind of logarithms being resummed, 
one might argue that there should only be one source of scale uncertainties.
Secondly, recall that the usual $\mu$ scale dependence arises 
because $\als$ depends on $\mu$ and we do not know its exact $\mu$ dependence.%
\footnote{In practice, there is also scale dependence from parton distribution functions (PDFs), 
but that is beside the point. One could talk about QED instead of QCD\@.}
That is why physical quantities like cross-sections depend on $\mu$ even though we demanded that they should not.
In contrast, being a Lorentz invariant quantity, $\als$ cannot depend on $\nu$, because
$\nu$ is a rapidity cutoff, i.e., a proxy for the frame-dependent boundaries separating collinear, soft, and anti-collinear modes.   
Then, the $\nu$-independence requirement for physical quantities would just render the perturbation series to be trivially independent of $\nu$ order-by-order in $\als$, 
leaving us no additional scale uncertainties.
Indeed, in \Refs{Becher:2010tm, Becher:2011pf, Becher:2011xn, Becher:2012qa}, 
all-order factorization formulae were derived that are explicitly $\nu$ independent, 
based on a form of analytic regulator~\cite{Becher:2011dz} for cutting off rapidity divergences.
However, in a seminal paper on this subject~\cite{Chiu:2012ir},
a thorough, explicit analysis demonstrates that the $\nu$-RG constitutes   
an independent source of scale uncertainties, where $\nu$ is varied independently of $\mu$. A recent work~\cite{Neill:2015roa} also shows that rapidity scale dependence can have significant numerical impact.

In this paper, we provide a simple resolution to this puzzle. 
We will show exactly how $\nu$ dependence is concealed in the seemingly $\nu$-independent factorization formula in \Ref{Becher:2012qa},
by deriving an alternative all-order factorization formula.
Our formula is equivalent to that of \Ref{Becher:2012qa} if we were to compute to all orders in perturbation theory.
They are not equivalent when truncated at a finite order, however, 
and our formula vividly shows how it acquires $\nu$ dependence at a finite order 
and allows us to quantify the scale uncertainty associated with the precise choice of $\nu$.
Our understanding also captures how $\nu$ dependence arises in spite of $\als$ being $\nu$ independent, as well as how it constitutes an independent source of scale uncertainties even though there is only one kind of logarithms being resummed.

For the derivation of our factorization formula,
we will employ a form of analytic regulator to regulate rapidity divergences
that is similar to, but different from, what has used by \Refs{Becher:2011dz, Becher:2012qa}.
Our analytic regulator shares the same great advantage
that there is no need to perform `zero-bin subtraction'~\cite{Manohar:2006nz} to avoid double-counting modes in the overlapping regions,
as the integrals in the overlapping regions become scaleless and thus vanish once consistently multipole-expanded~\cite{Becher:2013xia}.
This type of analytic regulator is also shown to preserve fundamental properties such as gauge invariance~\cite{Becher:2011dz}.
The difference between our and their forms of analytic regulator is that our regulator treats the collinear and anti-collinear sectors in a symmetric manner by introducing two independent rapidity RG scales, 
while 
the regulator of \Refs{Becher:2011dz, Becher:2012qa} treats the two sectors asymmetrically but needs only one rapidity scale. 
As a consistency check, 
we will show in \App{asymmetric} that our factorization formula can also be obtained from the regulator of \Refs{Becher:2011dz, Becher:2012qa}.
Therefore, we can continue to take advantage of the enormous technical simplicity and ease of the analytic regulator approach for practical calculations 
while correctly taking into account rapidity scale uncertainties at the same time.

As an application of our factorization formula and method of estimating rapidity scale uncertainty,
we will update our earlier work of the $WW$ production with jet-veto at the LHC \cite{Jaiswal:2014yba},
which used the formulation of \Ref{Becher:2012qa} and thereby lacked scale uncertainties from $\nu$ dependence as well as suffered from another accidental cancellation 
that led to an underestimation of the total scale uncertainties.
We will see that, while the central values of the predictions remain completely unaltered, 
the scale uncertainties increase and the consistency between the next-to-leading logarithm (NLL) and next-to-next-to-leading logarithm (NNLL) calculations improves.

\section{The set-up, ingredients, and goal}
Even though the conceptual points we will be making are quite general, 
for definiteness we discuss a SCET formalism that is applicable to the resummation of jet-veto logarithms for the production of a color-neutral object~\cite{Berger:2010xi}, such as a higgs boson~\cite{Berger:2010xi, Becher:2012qa, Liu:2012sz} \cite{Becher:2013xia} \cite{Shao:2013uba, Li:2014ria, Boughezal:2014qsa}, 
$W^+ W^-$~\cite{Jaiswal:2014yba, Becher:2014aya} or other di-bosons~\cite{Wang:2015mvz}, at a hadron collider.
Jet veto rejects an event if it contains a jet whose transverse momentum is larger than a prescribed jet-veto scale $\ptv$.%
\footnote{We consider the simplest case where $\ptv$ is constant, 
but it could in principle be chosen to depend on other variables in the process, 
such as the rapidity of the jet~\cite{Gangal:2014qda}.}
Using $M$ to denote the invariant mass of the color-neutral system,
and keeping only the relevant features explicit, 
we can express the differential jet-veto cross-section with respect to $M$ as%
\footnote{If the color-neutral state is a higgs boson, it would be a total jet-veto cross-section with $M$ fixed to the higgs boson mass.}
\beq
\fr{\dd \sg}{\dd M}
\sim 
H(\mu) \, \soft(\mu,\nu,\nubar) \, B(\mu,\nu) \, \Bbar(\mu, \nubar) 
\,,\lb{e.xsec}
\eeq
where $\mu$ is the usual RG scale associated with UV divergence, 
while both $\nu$ and $\nubar$ are rapidity RG scales.
For generality, we have introduced two rapidity scales 
because in principle the location of collinear-vs-soft boundary is independent of that of soft-vs-anti-collinear boundary.  
We now define and explain each of $H$, $\soft$, $B$ and $\Bbar$ below.

The \emph{hard function} $H(\mu)$ is the squared amplitude for the hard parton collision with the center-of-momentum energy $M$ that produces the color-neutral system in question. 
Being identical to the corresponding squared amplitude in the full theory, the hard function has no rapidity divergences and is hence independent of $\nu$ and $\nubar$.
The scale $\mu$ has the usual interpretation as the scale or boundary beyond which the modes would be regarded as necessarily highly off-shell.
So, $\mu$ is a proxy for the level of virtuality of the degrees of freedom in the EFT, which suggests $\mu^2 \sim p^2 \sim \bar{p}^2 \sim p_\mathrm{s}^2 \sim \la^2 M^2$.
Further noting that $p_\perp \sim \bar{p}_\perp \sim p_{\mathrm{s}\perp} \sim \la M \sim \ptv$ in the presence of jet-veto, 
we expect $\mu \sim \ptv$.
Later we will see how this choice is indeed forced upon us by the formalism.
The job of $\mu$-RGEs is thus to take us from the high scale $\mu \sim M$ (the \emph{hard scale}) where the SCET is matched onto the full theory, down to the low scale $\mu \sim \ptv$ (the \emph{factorization scale}) where the cross-section~\eq{xsec} is evaluated,
resumming some of the logarithms of the form $\log(M / \ptv)$. 
(The remaining logarithms of the form $\log(M / \ptv)$ will be resummed by
rapidity RGEs, as we discuss below.)

The \emph{beam functions} $B(\mu, \nu)$ and $\Bbar(\mu, \nubar)$~\cite{Stewart:2009yx, Stewart:2010qs}
are respectively the PDFs for the collinear and anti-collinear partons \emph{with the jet-veto.}
That is, if we schematically write a quark PDF as
\beq
\phi \sim \sum_X \bra N| \wba{\psi} |X\ket \bra X| \psi |N\ket
\,,
\eeq
where $\psi$ and $N$ are respectively the collinear quark field and the nucleon in question, and the summation goes over all possible hadronic states $X$,
then the beam function for the same quark and nucleon is given by  
\beq
B \sim \sum_X\raise.5em\hbox{\!$'$} \bra N| \wba{\psi} |X\ket \bra X| \psi |N\ket
\,,
\eeq
which is the same as before down to every little (implicit) detail except that the summation $\sum'$ only goes over $X$ that satisfies the jet-veto condition.
The collinear beam function $B(\mu, \nu)$ suffers from 
rapidity divergence when we send $k_+$ (the $+$ component of integration variables) toward zero and 
thus `invade' the territory of soft modes. 
The regulator we introduce in \Sec{derivation} regulates this divergence and 
trades it for a scale $\nu$.
This regulator does not introduce any other new scale, 
so we expect the logarithms from $k_+$ integration to have the form $\log (\nu / \xi P_+)$,
because the only physical scale in the $+$ direction is  
$\xi P_+$ (where $\xi$ is the parton momentum fraction and $P_+/2$ the proton momentum in the collinear beam). 
This suggests that the good scale choice that minimizes the logarithms should be $\nu \sim \xi P_+$.
Similarly, the anti-collinear beam function $\Bbar(\mu, \nubar)$ has rapidity divergence as $k_- \to 0$, which is regulated in favor of a scale $\nubar$.
Our regulator by design treats the collinear and anti-collinear sectors symmetrically, 
so we expect $\nubar \sim \xibar P_-$ (where $\xibar$ and $P_-/2$ are the parton momentum fraction and proton momentum in the anti-collinear beam).
Integrations over the $\perp$ component are also divergent, 
which we regulate by dimensional regularization acting on the $d-2$ transverse dimensions with the scale $\mu$.
As we already discussed, the $\perp$ component controls the degree of virtuality of the partons, so we expect the $\mu$-dependent logarithms to have the form $\log(\mu / \ptv)$ in the presence of jet-veto, suggesting the choice $\mu \sim \ptv$.

Finally, $\soft(\mu, \nu, \nubar)$ is called the \emph{soft function}, 
which is a matrix element of a SCET operator consisting only of soft modes, 
evaluated between states containing only soft modes.
Rapidity divergences arise when we send $k_+$ or $k_-$ to infinity and invade the land of collinear or anti-collinear modes. 
The regulator trades the $k_+$ and $k_-$ divergences for $\nu$ and $\nubar$, respectively.
Alternatively, in our context, the expression~\eq{xsec} tells us that $\soft$ can also be regarded as a renormalization constant that absorbs the rapidity divergences and $\nu$- and $\nubar$-dependences of the product of operators $B \Bbar$.
In the former interpretation, the soft modes are `integrated in' in the SCET
and we can directly calculate the soft function diagrammatically.
In the latter, the soft modes are not in the theory 
and $\soft$ is determined by matching $\soft B \Bbar$ onto the full theory at some appropriate scale. 
Either way, what we actually do in practice are essentially identical,
while both interpretations can be useful conceptually. 
So we will discuss the two viewpoints in parallel below.
Most importantly, whether it is viewed as a soft function or renormalization constant, 
$\soft$ should not know anything about $\xi P_+$, $\xibar P_-$, nor $M$. 
Thus, the degree of virtuality imposed by jet-veto, $\mu \sim \ptv$, 
is the only physical scale entering $\soft$, so the good scale choice must be given by $\nu \sim \nubar \sim \mu$ to avoid large $\log(\nu / \mu)$ and $\log(\nubar / \mu)$.

We can now state our goal.
For $B$ and $\Bbar$, 
we have high scales $\nu \sim \xi P_+$ and $\nubar \sim \xibar P_-$,
which are both of order $M$.
For $\soft$, we have low scales $\nu \sim \nubar \sim \mu$,
which is of order $\ptv$. 
The job of rapidity RGEs is then to take us from the high scales to the low scales, 
resumming the logarithms of the form $\log(M / \ptv)$,
the same form as what the $\mu$-RGEs resum.
As we alluded in \Sec{intro},
this is not surprising 
as there is actually only one large ratio of scales, $1/\la$, in the theory.
This is also reflected to the correlation between the rapidity scales and the virtuality scale, $\nu \sim \nubar \sim \mu$.
So, our problem is to analyze how imperfect this correlation can be at a finite order in perturbation theory and to quantify the associated uncertainty.

\section{(Rapidity) RG equations}
\lb{s.derivation}
In this section, we will derive an all-order form of factorization formula to be used in \Eq{xsec}, that is, the precise form of ``$\soft(\mu,\nu,\nubar) \, B(\mu,\nu) \, \Bbar(\mu, \nubar)$'' there.
The reader who wishes to skip the derivation and read about the discussion of rapidity scale uncertainty may directly move on to \Sec{results}\@. 

To regulate rapidity divergences,
we employ the following form of analytic regulator,
where for each phase-space integral%
\footnote{As shown in \Ref{Becher:2011dz}, rapidity divergences only occur in phase space integrals.}
with momentum $k$ we insert   
\beq
 \lt( \fr{\nu}{k_+}    \rt)^{\!\! \al}    \theta(k_+ - k_-) 
 + 
 \lt( \fr{\nubar}{k_-} \rt)^{\!\! \albar} \theta(k_- - k_+) 
\,.\lb{e.ana_reg}
\eeq 
For a collinear $k$ and at the leading order in $\lambda$ (i.e., $\co(\la^0)$), 
we should neglect $k_-$ in comparison with $k_+$ inside $\theta(k_+ - k_-)$,
which renders the step function trivial.  
Thus, this regulator in practice simply amounts to inserting $( \nu / k_+ )^\al$ in the phase-space integration for each collinear particle,
and similarly just inserting $( \nubar / k_- )^\albar$ for each anti-collinear particle.
This symmetric treatment of the collinear and anti-collinear sectors then implies that 
$B$ and $\Bbar$ are identical up to the trivial relabelling $\nu \lr \nubar$, $\al \lr\albar$, etc.

As all regulators do, the regulator~\eq{ana_reg} introduces artificial new scales  into the problem, $\nu$ and $\nubar$, which are proxies for the boundaries separating collinear modes from soft modes, and anti-collinear modes from soft modes, respectively. 
The requirement that physical observables should be independent of $\nu$ and $\nubar$ leads to rapidity RGEs.
We also introduce separate jet-veto scales $\ptv$ and $\ptvbar$ for the collinear and anti-collinear sectors, respectively, as they can in principle be chosen independently and it has proven useful to do so~\cite{Becher:2012qa}.%
\footnote{However, we still assume $\ptvbar \sim \ptv$ to avoid introducing another hierarchy of scales into the theory.
We will see later that the formalism would indeed break down if $\ptvbar$ is very different from $\ptv$.}
Now, as we take the $\al \to 0$ and $\albar \to 0$ limits, 
$1 / \al$ and $1 / \albar$ poles appear.
Just like a $1 / \ep$ pole in dimensional regularization comes with $\log\mu$, 
the $1 / \al$ and $1 / \albar$ poles are respectively accompanied by
\beq
L \equiv \log \fr{\nu}{\xi P_+}
\,,\quad
\Lbar \equiv \log \fr{\nubar}{\xibar P_-}
\,.\lb{e.L_and_Lbar}
\eeq  
Here, $\xi P_+/2$ and $\xibar P_-/2$ are respectively the energies of the collinear and anti-collinear partons that enter the hard function $\hard$,
while $P_+/2$ and $P_-/2$ are the energies of their respective parent hadrons.
These logarithms are the only sources of $\nu$ and $\nubar$ dependences.
The form of $L$ can be understood by observing that the only physical scale entering in $k_+$ integration in the collinear sector is $\xi P_+$. 
The form of $\Lbar$ then follows from symmetry.

The beam functions also depend on the scale $\mu$ of dimensional regularization,
which regulates divergence from $k_\perp$ integration.
Then, since the jet-veto scale is the only physical scale in the $\perp$ direction in the EFT (as the hard physics of order $M$ has been already integrated out), 
the only dependence on $\mu$ besides through $\als(\mu)$ is via the following logarithms: 
\beq
\Lp    \equiv \log \fr{\mu^2}{(\ptv)^2}    
,\quad
\Lpbar \equiv \log \fr{\mu^2}{(\ptvbar)^2} 
.\lb{e.Lp_and_Lpbar}
\eeq
%

\subsection{The $\mu$-RG equations}
\lb{s.mu_RG}
Without loss of generality, 
the $\mu$-RGEs can be written as
\beq
\mu \fr{\del}{\del\mu} \log B     (\mu, \nu)
&=  
\sum_{p=0}^\infty f^{(p)\!} (\xi,    \als, \Lp) \,    L^p
\,,\\
\mu \fr{\del}{\del\mu} \log \Bbar (\mu, \nubar) 
&=  
\sum_{p=0}^\infty f^{(p)\!} (\xibar, \als, \Lpbar) \, \Lbar^p
\,,\lb{e.mu_RG.naive}
\eeq
where the $\mu$ dependences are in $\als$, $\Lp$, and $\Lpbar$, 
while the $\nu$ and $\nubar$ dependences are in $L$ and $\Lbar$, respectively.
The functions $f^{(p)}$ in the $\Bbar$ equation are the same functions as those in the $B$ equation, owing to the symmetry of analytic regulator~\eq{ana_reg}.

The equations~\eq{mu_RG.naive} are actually too general and their forms can be constrained significantly. 
Physical observables of our interest, such as the $WW$ jet-veto cross-section, 
depend only on the product $\soft B \Bbar$, as in \Eq{xsec}.
Being a renormalization constant, 
$\soft$ cannot depend on dynamical variables such as $\xi$, $\xibar$, $P_+$, or $P_-$.
Alternatively, being the soft function, i.e., 
a matrix element of soft fields between soft states, 
it cannot depend on the momenta of the (anti-)collinear partons nor of the colliding hadrons.
We therefore require that the sum of the right-hand sides of the $\mu$-RGEs~\eq{mu_RG.naive}, i.e., 
\beq
A \equiv 
  \sum_{p=0}^\infty f^{(p)\!} (\xi,    \als, \Lp)    \, L^p  
+ \sum_{p=0}^\infty f^{(p)\!} (\xibar, \als, \Lpbar) \, \Lbar^p
\,,\lb{e.A} 
\eeq 
should be independent of $\xi$, $\xibar$, $P_+$, and $P_-$ at any values of 
$\mu$, $\nu$, and $\nubar$.
These four variables, however, are not all independent of each other 
but are constrained as $\xi P_+ \xibar P_- = M^2$
because we are considering the quantity of the form~\eq{xsec},
which is a function of $M$.
So, we instead consider
\beq
\wtd{A} \equiv 
A + \ell \lt( \fr{\xi P_+ \xibar P_-}{M^2} - 1 \rt)\!
\,,
\eeq
where $\ell$ is a Lagrange multiplier varied independently of $\xi$, $\xibar$, $P_+$, and $P_-$, so that the condition~$\del \wtd{A} / \del \ell = 0$ gives us the 
constraint back.
Then, the requirement~$\del\wtd{A} / \del P_+ =0$ implies
\beq
f^{(1)} = \ell 
\,,\quad
f^{(2)} = f^{(3)} = \cdots = 0 
\,.\lb{e.condition.1}
\eeq
The first relation here means that $f^{(1)\!}(\xi, \als, \Lp)$ is actually independent of $\xi$, 
i.e., $f^{(1)\!}(\xi, \als, \Lp) = f^{(1)\!}(\als, \Lp)$,
because the Lagrange multiplier $\ell$ is by definition independent of the remaining independent variables, $\xi$, $\xibar$, $P_+$, and $P_-$.
However, $\ell$ may still depend on other parameters and variables. 
Most generally, it may depend on $\als$.
It cannot actually depend on $\Lp$ nor $\Lpbar$ for the following reason.
Our symmetric treatment of the collinear and anti-collinear sectors implies that if $\ell$ depended on $\Lp$, it should also depend on $\Lpbar$ symmetrically, i.e., $\ell(\als, \Lp, \Lpbar) = \ell(\als, \Lpbar, \Lp)$.
However, the above condition $f^{(1)\!}(\als, \Lp) = \ell$ tells us that $\ell$ does not depend on $\Lpbar$, so by symmetry it does not depend on $\Lp$ either. 
We thus conclude that $f^{(1)\!}(\als, \Lp) = f^{(1)\!}(\als)$.
Then, applying this and the conditions~\eq{condition.1} to the requirement~$\del\wtd{A} / \del \xi = 0$, we get
\beq
\fr{\del}{\del \xi} f^{(0)\!}(\xi, \als, \Lp) = 0
\,.
\eeq
We thus have narrowed down the $\mu$-RGEs~\eq{mu_RG.naive} to
\beq
\mu \fr{\del}{\del\mu} \log B     (\mu, \nu)
&=  
f^{(1)\!}(\als) \,  L    + f^{(0)\!} (\als, \Lp) 
\,,\\
\mu \fr{\del}{\del\mu} \log \Bbar (\mu, \nubar)
&=  
f^{(1)\!}(\als) \, \Lbar + f^{(0)\!} (\als, \Lpbar) 
\,.\lb{e.mu_RG}
\eeq

Next, since the hard function $\hard(\mu)$ is independent of $\nu$ and $\nubar$, 
the $\nu$ and $\nubar$ dependences in the product $B\Bbar$ should be completely cancelled by $\soft$ in the product $\soft B \Bbar$.
So, we must have
\beq
\mu \fr{\del}{\del\mu} \log \soft (\mu, \nu, \nubar)
= -f^{(1)\!}(\als) \, \Ls + F^{(0)\!}(\als, \Lp, \Lpbar)
\,,\lb{e.mu_RG.soft}  
\eeq
where
\beq
\Ls \equiv \log \fr{\nu \nubar}{\mu^2}
\,.
\eeq

We can determine $f^{(1)}$ and a combination of $f^{(0)}$ and $F^{(0)}$ from 
the requirement that the $\mu$ dependence in the product $\soft B \Bbar$ must be cancelled when $\soft B \Bbar$ is multiplied by the hard function $\hard$,
i.e., $\mu \, \del \log(\soft B \Bbar \hard) / \del \mu = 0$.
The $\mu$-RGE of the hard function can be parametrized as
\beq
\mu \fr{\dd}{\dd\mu} \log \hard(\mu) = 2\Ga \LM + 4\ga
\,,\lb{e.mu_RG.hard}
\eeq
where $\LM \equiv \log \!\lt( M^2 / \mu^2 \rt)$ 
with the coefficients $\Ga$ (the cusp anomalous dimension) and $\ga$ (the anomalous dimension) that
can be calculated order-by-order in $\als$ using perturbation theory.
To identify $f^{(1)}$, we demand that the derivatives $\mu \, \del / \del \mu$ 
and $M \, \del / \del M$ commute when acting on $\log(\soft B \Bbar \hard)$.
This identifies the $M$ dependence in $\LM$ with that in $L + \Lbar$ from \Eq{mu_RG}, 
i.e.,%
\footnote{Mathematically, a $\mu$-independent integration constant should be included in the right-hand side of the relation~\eq{f1}.
However, such integration constant is absent in perturbation theory as all terms in RGEs contain $\als$, which does depend on $\mu$.}
\beq
f^{(1)\!}(\als) = 2\Ga 
\lb{e.f1}
\,.
\eeq
Cancelling the remaining $\mu$ dependence in $\log(\soft B \Bbar \hard)$, we  
then get
\beq
f^{(0)\!}(\als, \Lp) + f^{(0)\!}(\als, \Lpbar) + F^{(0)\!}(\als, \Lp, \Lpbar) = -4 \ga
\,.\lb{e.f0}
\eeq
%

\subsection{The $\nu$-RG equations}
\lb{s.nu_RG}
Physical quantities must be independent of the renormalization scheme.
In particular, the solution of RGEs must be independent of the shape of the path in the $\mu$-$\nu$-$\nubar$ space and can only depend on the endpoints of the path.
This path independence requires that
\beq
\mu \fr{\del}{\del\mu} \nu \fr{\del}{\del\nu} \log B (\mu, \nu)
&=
\nu \fr{\del}{\del\nu} \mu \fr{\del}{\del\mu} \log B (\mu, \nu)
\\
&= f^{(1)\!}(\als)
\,,\lb{e.path_independence.B}
\eeq
where the second equality has used \Eq{mu_RG}.
Thus, $\mu \, \del / \del\mu$ acting on $\nu \, \del \log B / \del \nu$ gives us 
a $\nu$-independent function, $f^{(1)\!}(\als)$.
So, if there are any $\nu$-dependent terms in $\nu \, \del \log B / \del \nu$, they must be \emph{in}dependent of $\mu$ so that they get annihilated by $\mu \, \del / \del\mu$.
However, since the perturbation series is an expansion in $\als$, 
every term does depend on $\mu$ via $\als$.
Therefore, there cannot be any $\nu$-dependent terms in $\nu \, \del \log B / \del \nu$. 
We thus arrive at the general form of rapidity RGEs:
\beq
\nu    \fr{\del}{\del\nu}    \log B     (\mu, \nu)
&=  
g(\mu) 
\,,\\
\nubar \fr{\del}{\del\nubar} \log \Bbar (\mu, \nubar) 
&=  
\gbar(\mu) 
\,,\lb{e.nu_RG.beam}
\eeq
where
\beq
g(\mu) \equiv \sum_{p=0}^\infty g^{(p)\!} (\als) \, \Lp^p  
\,,\quad
\gbar(\mu) \equiv \sum_{p=0}^\infty g^{(p)\!} (\als) \, \Lpbar^p
\,.\lb{e.g_gbar}
\eeq
(Note that these are expansions in $\Lp$ and $\Lpbar$ unlike the $\mu$-RGEs~\eq{mu_RG.naive}, which were expansions in $L$ or $\Lbar$.)
We have written $g^{(p)}$ as a function of only $\als$, 
as it turns out that $g^{(p)}$ is independent of $\xi$ (or $\xibar$) as we will see shortly.
We again require that the product $\soft B \Bbar$ to be independent of $\nu$ and $\nubar$, 
which implies that the rapidity RGEs for $\soft$ must be given by
\beq
\nu    \fr{\del}{\del\nu}    \log \soft (\mu, \nu, \nubar)
&=  
-g(\mu)  
\,,\\
\nubar \fr{\del}{\del\nubar} \log \soft (\mu, \nu, \nubar) 
&=  
-\gbar(\mu)
\,.\lb{e.nu_RG.soft}
\eeq
The path independence for $\soft$ does not provide us with any new constraints.
First, the $\soft$ rapidity RGEs~\eq{nu_RG.soft} trivially lead to $\lt[ \nu \, \del / \del\nu, \, \nubar \, \del / \del\nubar \rt] \log \soft = 0$.
Also, 
the path independence condition~\eq{path_independence.B} for $B$ and the similar condition for $\Bbar$ readily imply that 
$\lt[ \mu \, \del / \del\mu, \, \nu \, \del / \del\nu \rt] \log\soft 
= \lt[ \mu \, \del / \del\mu, \, \nubar \, \del / \del\nubar \rt] \log\soft 
= 0$. 

\subsection{The all-order factorization formula and the choice of scales}
\lb{s.factorization}
We are now ready to solve the rapidity RGEs~\eq{nu_RG.beam} and~\eq{nu_RG.soft} for an arbitrary but fixed $\mu$.
Solving them is trivial as the right-hand sides of all the $\nu$-RGEs are independent of $\nu$ and $\nubar$.
We solve for $B$ starting from an `initial' point $\nu = \nuinitial$ to an arbitrary final $\nu$,
$\Bbar$ from $\nubarinitial$ to $\nubar$, 
and $\soft$ from $(\nusinitial, \nusbarinitial)$ to $(\nu, \nubar)$.
Then, combining all the solutions, we get 
\beq
& \soft(\mu, \nu, \nubar) \, B(\mu, \nu) \, \Bbar(\mu, \nubar)
\\
&= 
\!\lt( \fr{\nusinitial}{\nuinitial} \rt)^{\! g(\mu)} \!
\!\lt( \fr{\nusbarinitial}{\nubarinitial} \rt)^{\! \gbar(\mu)} \!
\hat{\soft} \hat{B} \hat{\Bbar} (\mu)
\lb{e.factorization.original}
\eeq
with
\beq
\hat{\soft} \hat{B} \hat{\Bbar} (\mu) 
\equiv 
\soft(\mu, \nusinitial, \nusbarinitial) \, 
B(\mu, \nuinitial) \, \Bbar(\mu, \nubarinitial)
\,.\lb{e.hatted}
\eeq

To determine what value of $\mu$ we should choose when we evaluate 
$\soft(\mu, \nu, \nubar) \, B(\mu, \nu) \, \Bbar(\mu, \nubar)$ 
in the factorization formula~\eq{factorization.original},
let us examine more closely the path independence condition~\eq{path_independence.B}.
Applying the relation~\eq{f1} to it, we can rewrite it as a condition for $g(\mu)$:
\beq
\mu \fr{\dd}{\dd\mu} g(\mu) = f^{(1)\!}(\als) = 2\Ga
\,.\lb{e.path_independence}
\eeq
Satisfying this condition order-by-order in $\Lp$ in the expansion~\eq{g_gbar}, 
we get the following nontrivial recursion relations among the coefficient functions $g^{(p)\!}(\als)$:
\beq
2  g^{(1)} + \be g'^{(0)}     &= f^{(1)} = 2\Gamma
\,,\\
2p g^{(p)} + \be g'^{(p-1)} &= 0 
\qquad (p \geq 2)
\,,\lb{e.recursion}
\eeq 
where $g'^{(p)} \equiv \dd g^{(p)} / \dd \als$ 
and $\be \equiv \mu \, \dd \als / \dd \mu$.
Let us parametrize $g^{(0)}$, $\Ga$, and $\be$ as
\beq
g^{(0)}   &= \as \, d_1  + \as^2 \, d_2 + \cdots 
\,,\\
\Ga   &= \as \!\lt( \Ga_0 + \as \Ga_1 + \cdots \rt)
,\\
\be &= - 2 \als \as \!\lt(  \be_0 + \as \be_1 + \cdots \rt)
,\lb{e.parametrization}
\eeq
where $\as \equiv \als / (4\pi)$, and
all the coefficients, $d_{1,2,\cdots}$, $\Ga_{0,1,\cdots}$, $\be_{0,1,\cdots}$, 
can be calculated in perturbation theory.
Since none of these coefficients depend on $\xi$, 
the functions $g^{(p)}$ are indeed independent of $\xi$ for all $p$ 
as we alluded earlier.
Combined with the parametrization~\eq{parametrization}, 
the recursion relations~\eq{recursion} give
\beqa{5}
g^{(1)} &= \as \Ga_0 \,+\, && \as^2 (\Ga_1 + \be_0 d_1) && + \cdots 
\,,\\
g^{(2)} &=                 && \as^2 \fr{\Ga_0 \be_0}{2} && + \cdots
\,.\lb{e.g1g2}
\eeqa
Note that $g^{(p)} = \co(\als^p)$ for all $p \geq 1$.
This means that in order for us to be able to reliably truncate 
the expansions~\eq{g_gbar}
at any finite order in $\als$, 
the logarithms $\Lp$ and $\Lpbar$ must be small, i.e., $\Lp, \Lpbar \ll \als^{-1}$.
This is possible only if we choose 
\beq
\mu \sim \ptv \sim \ptvbar
\,,\lb{e.mu_choice}
\eeq
There are two things to note here.
Firstly, 
the right choice of the factorization scale at which we should evaluate
the factorization formula~\eq{factorization.original} is on the order of the jet-veto scale $\ptv$, rather than the hard scale $M$.
Secondly, our formalism would not be applicable to the case in which $\ptv$ and $\ptvbar$ are hierarchically different.
In such case, there would remain large logarithms of the form $\log (\ptv / \ptvbar)$ that are not resummed by our formalism. 

The remaining scales $\nuinitial$, $\nubarinitial$, $\nusinitial$, and $\nusbarinitial$ should be chosen to minimize $L$, $\Lbar$, and $\Ls$.
These logarithms vanish if
\beq
\nuinitial    = \xi P_+
\,,\quad
\nubarinitial = \xibar P_-
\,,\quad
\nusinitial = \nusbarinitial = \mu 
\,.\lb{e.nu_choice}
\eeq
The factorization formula~\eq{factorization.original} then becomes
\beq
& \soft(\mu, \nu, \nubar) \, B(\mu, \nu) \, \Bbar(\mu, \nubar)
\\
&= 
\!\lt( \fr{\mu}{\xi P_+}    \rt)^{\! g(\mu)} \!
\!\lt( \fr{\mu}{\xibar P_-} \rt)^{\! \gbar(\mu)} \!
\hat{\soft} \hat{B} \hat{\Bbar} (\mu)
\,.\lb{e.factorization.full}
\eeq
However, since our formalism is valid only for $\ptvbar \sim \ptv$, 
there is no practical benefit to be gained by choosing the two jet-veto scales
differently in experimental analyses. 
So, hereafter, we will mostly focus on the special case, 
\beq
\ptvbar = \ptv
\,,\quad
\Lpbar = \Lp
\,,\quad
\gbar = g
\,.
\eeq
The factorization formula~\eq{factorization.full} now takes a very simple form:
\beq
\soft(\mu, \nu, \nubar) \, B(\mu, \nu) \, \Bbar(\mu, \nubar)
= 
\!\lt(\! \fr{\mu^2}{M^2} \!\rt)^{\!\! g(\mu)} \!
\hat{\soft} \hat{B} \hat{\Bbar} (\mu)
\,.\lb{e.factorization.exact}
\eeq

Before closing this section, 
we would like to remind the reader that the only properties of the analytic regulator~\eq{ana_reg} that were actually used in 
the derivation of the formula~\eq{factorization.exact} 
are the forms of the logarithms~\eq{L_and_Lbar} and~\eq{Lp_and_Lpbar}.
Therefore, any other regulator that only produces those logarithms will lead to 
exactly the same form of factorization formula as \Eq{factorization.exact}. 
It is actually even more robust than that 
as we will see in \App{asymmetric} where 
a regulator that gives different forms of logarithms still arrives at 
essentially the same factorization formula,
although it seems difficult to pin down exactly how robust it is.
Finally, let us also remark that it should be straightforward to repeat our analysis for objects other than the jet-veto beam and soft functions, 
such as the transverse momentum dependent parton distribution functions (TMDPDFs)
for small $\pT$ resummation.

\section{Main results and discussions}
\lb{s.results}
Let us now examine what happens to the factorization formula~\eq{factorization.full} 
or \eq{factorization.exact} if we deviate from the scale choice~\eq{nu_choice}.
We are now ready to show that, 
while such deviations cancel out order-by-order in $\als$, 
the cancellation is incomplete if the series is truncated in a way consistent with the EFT power-counting rules, rendering rapidity RG to be an independent source of scale uncertainty.

\subsection{Finite-order truncation and additional scale uncertainties from rapidity RG\@.}
\lb{s.truncation}
If we could calculate $\soft B \Bbar$ exactly or to all orders in perturbation theory,
the scale choice~\eq{nu_choice} would be unique and unambiguous. 
This is a corollary of the fact that the right-hand sides of $\nu$-RGEs~\eq{nu_RG.beam} and~\eq{nu_RG.soft} are independent of $\nu$ and $\nubar$.
For example, one might object to the choice~\eq{nu_choice} that the vanishing of $\Ls$ only implies $\nusinitial \nusbarinitial = \mu^2$, 
so we could instead choose $\nusinitial = r \mu$ and $\nusbarinitial = r^{-1} \mu$ with some $r$. 
Glancing back at \Eq{factorization.original}, 
we see that this (apparently) alternative choice would amount to 
multiplying the prefactor $(\mu / \xi P_+)^g \, (\mu / \xibar P_-)^\gbar$ in the factorization formula~\eq{factorization.full} by a factor of $r^{g - \gbar}$,
while changing $\hat{\soft}$ to $\soft(\mu, r\mu, r^{-1}\mu)$ at the same time.
But the $\nu$-RGE~\eq{nu_RG.soft} tells us that $\soft(\mu, r\mu, r^{-1}\mu) = r^{-g+\gbar} \hat{\soft}$, 
so $r$ cancels out. 
In fact, any one of $\nuinitial$, $\nubarinitial$, $\nusinitial$ and $\nusbarinitial$ can be varied independently, e.g., $\nuinitial =\xi P_+ \to r \, \xi P_+$, 
without producing any net effect.
This `$r$ invariance' is exact, 
and since $\als$ is independent of $\nu$ and $\nubar$, 
it is exact order-by-order in $\als$.

However, the whole point of using an EFT is to resum the large logarithm $\log(1/\la)$.
So, in the EFT power counting, $\als \log(1/\la)$ is by definition parametrically regarded as $\co(1)$.
Thus, the EFT perturbation series is \emph{not} a literal power series in $\als$ 
but an effective power series in $\als$ \emph{with $\log(1/\la)$ counted as $\co(\als^{-1})$.}
When the factorization formula~\eq{factorization.exact} is truncated at a finite order according to this EFT power counting, the $r$ invariance is no longer exact.
Let us say we are aiming at an $\co(\als^k)$ accuracy
with the EFT power counting (i.e., the so-called N$^{k+1}$LL accuracy).
Then, the right-hand side of factorization formula~\eq{factorization.exact} 
should be truncated as
\beq
\!\lt(\! \fr{\mu^2}{M^2} \!\rt)^{\!\! h_1 \als + h_2 \als^2 + \cdots + h_{k+1} \als^{k+1}} \!
\!\Bigl[ \hat{\soft} \hat{B} \hat{\Bbar} (\mu) \Bigr]_\text{computed to $\co(\als^k)$}
\,,\lb{e.central}
\eeq
where $h_n$ are $\als$-independent coefficients defined via 
the expansion $g(\mu) = \sum_{n=1}^\infty h_n \, \als^n$. 
Recalling the definition~\eq{hatted} and the scale choices~\eq{mu_choice} and~\eq{nu_choice}, 
we see that the object $\bigl[ \hat{\soft} \hat{B} \hat{\Bbar} (\mu) \bigr]$ 
has no logarithms of $\nu$ or $\nubar$ at all, nor any large logarithms of $\mu$.
That is why the object $\bigl[ \hat{\soft} \hat{B} \hat{\Bbar} (\mu) \bigr]$ is calculated above to literally $\co(\als^k)$, 
with no need to take into account the EFT power counting.
In contrast, the exponent of $\mu^2 / M^2$ is computed above to $\co(\als^{k+1})$, 
because $\log(M^2 / \mu^2) \sim \log(M^2 / (\ptv)^2) \sim \log(1/\la) \sim \co(\als^{-1})$ and the EFT power counting must be applied. 

Let us now examine how $\soft B \Bbar$ changes from the `central value'~\eq{central} if we deviate from the scale choice~\eq{nu_choice}. 
Again, looking back at \Eq{factorization.original} (with $\gbar = g$) 
and using the $\nu$-RGEs~\eq{nu_RG.beam} and~\eq{nu_RG.soft}, 
we see that the prefactor $(\mu^2 / M^2)^{h_1 \als + h_2 \als^2 + \cdots + h_{k+1} \als^{k+1}}$ will get multiplied by $r^{h_1 \als + h_2 \als^2 + \cdots + h_{k+1} \als^{k+1}}$.
For the product of matrix elements $\bigl[ \hat{\soft} \hat{B} \hat{\Bbar} (\mu) \bigr]_\text{to $\co(\als^k)$}$, there are two possibilities depending on how the $r$ dependence is calculated, which agree to $\co(\als^k)$ but differ at $\co(\als^{k+1})$ and hence leading to different estimates of rapidity scale uncertainty of $\co(\als^{k+1})$. 
One way is to start with an $\co(\als^k)$-truncated ``initial'' $\hat{\soft} \hat{B} \hat{\Bbar}$ evaluated at the scale~\eq{nu_choice} and then evolve it to a different scale choice by using the $\co(\als^k)$-truncated $\nu$-RGEs.
The other way is to directly evaluate $\hat{\soft} \hat{B} \hat{\Bbar}$ at $\co(\als^k)$ at a point away from the choice~\eq{nu_choice}.   
In the first method, 
the product of matrix element $\bigl[ \hat{\soft} \hat{B} \hat{\Bbar} (\mu) \bigr]_\text{to $\co(\als^k)$}$ would be simply multiplied by
$r^{-h_1 \als - h_2 \als^2 - \cdots - h_k \als^k}$.
Then, combined with the factor of $r^{h_1 \als + h_2 \als^2 + \cdots + h_{k+1} \als^{k+1}}$ coming from $(\mu^2 / M^2)^{h_1 \als + h_2 \als^2 + \cdots + h_{k+1} \als^{k+1}}$, we would be left with a simple net $r$ dependence, $r^{h_{k+1} \als^{k+1}}$.
This is not optimal for the purpose of estimating rapidity scale uncertainty, however,
because it is only sensitive to a single coefficient $h_{k+1}$, which may be accidentally small or large.
On the other hand, the second method leads to
\beq
& \soft(\mu, \nu, \nubar) \, B(\mu, \nu) \, \Bbar(\mu, \nubar)
\\
&=
\!\lt(\! \fr{\mu^2}{M^2} \!\rt)^{\!\! h_1 \als + \cdots + h_{k+1} \als^{k+1}} \!
r^{h_1 \als + h_2 \als^2 + \cdots + h_{k+1} \als^{k+1}} \\
&\quad\times
\bigl[ r^{-h_1 \als - h_2 \als^2 - \cdots - h_k \als^k}
\hat{\soft} \hat{B} \hat{\Bbar} (\mu) \bigr]_\text{to $\co(\als^k)$}
\,,\lb{e.factorization.expanded}
\eeq
which is more robust as it involves not only all of $h_1$ through $h_{k+1}$ 
but also the whole perturbative series of $\hat{\soft} \hat{B} \hat{\Bbar} (\mu)$ to $\co(\als^k)$.  
Hence, we expect that varying $r$ by an $\co(1)$ factor in the expression~\eq{factorization.expanded} should give us a plausible estimate of rapidity scale uncertainty of $\co(\als^{k+1})$.
We therefore propose the formula~\eq{factorization.expanded} as our N$^{k+1}$LL truncation of the factorization formula~\eq{factorization.exact} with rapidity scale uncertainty appropriately taken into account.

\subsection{Comparison with the literature}
We can now discuss exactly what is missing in the all-order factorization formula of \Ref{Becher:2012qa}, which exhibits no rapidity scale dependence.
They used an analytic regulator different from \Eq{ana_reg},
which treats the collinear and anti-collinear sectors differently, 
but as we will show in \App{asymmetric}, it nonetheless leads to 
a factorization formula essentially identical to \Eq{factorization.exact}: 
\beq
\soft(\mu) \, B(\mu, \nu) \, \Bbar(\mu, \nu)
= 
\!\lt(\! \fr{\mu^2}{M^2} \!\rt)^{\!\! g(\mu)} \!
\hat{\soft} \hat{B} \hat{\Bbar} (\mu)
\,,\lb{e.factorization.asymm}
\eeq
where $\hat{\soft} \hat{B} \hat{\Bbar} (\mu) 
\equiv 
\soft(\mu) \, 
B(\mu, \nuinitial) \, \Bbar(\mu, \nubarinitial)$
and we have taken $\ptvbar = \ptv$ as we did for \Eq{factorization.exact}. 
Then, repeating the analysis of \Sec{truncation},  
we arrive at a correct N$^{k+1}$LL expression
\beq
& \soft(\mu) \, B(\mu, \nu) \, \Bbar(\mu, \nu)
\\
&=
r^{h_{k+1} \als^{k+1}} 
\!\lt(\! \fr{\mu^2}{M^2} \!\rt)^{\!\! h_1 \als + \cdots + h_{k+1} \als^{k+1}} \!
\!\Bigl[ \hat{\soft} \hat{B} \hat{\Bbar} (\mu) \Bigr]_\text{to $\co(\als^k)$}
\,,\lb{e.factorization.expanded.asymm}
\eeq
which is essentially the same as~\Eq{factorization.expanded}.

To obtain the form of factorization formula of \Ref{Becher:2012qa} from \Eq{factorization.asymm},
we first set $r=1$, since \Ref{Becher:2012qa} has made the scale choice $\nuinitial = \xi P_+$ and $\nubarinitial = \mu^2 / (\xibar P_-)$ (see \App{asymmetric}) implicitly and did not consider variations from it.
We also rewrite $\mu^2 / M^2$ as  $\mu^2 / M^2 = [(\ptv)^2 / M^2] \, \e^{\Lp}$.
Then, instead of the formula~\eq{factorization.expanded.asymm}, we would get
\beq
& \soft(\mu) \, B(\mu, \nu) \, \Bbar(\mu, \nu)
\\
&=
\!\lt(\! \fr{(\ptv)^2}{M^2} \!\rt)^{\!\! h_1 \als + \cdots + h_{k+1} \als^{k+1}} \!
\!\Bigl[ \soft' B' \Bbar' (\mu) \Bigr]_\text{to $\co(\als^k)$}
\,,\lb{e.naive}
\eeq
where $\soft' B' \Bbar' (\mu) 
\equiv
\e^{g \Lp} \hat{\soft} \hat{B} \hat{\Bbar} (\mu)$. 
This is the form presented in \Ref{Becher:2012qa}.
Since there is no $r$ dependence, there is no rapidity scale uncertainty to talk about.
One might argue that the lack of the $r$-dependent prefactor $r^{h_{k+1} \als^{k+1}}$ could be (retrospectively) justified by saying that it is $\co(\als^{k+1})$ for $r = \co(1)$.
However, in our analysis of \Sec{truncation},
the factor $r^{h_{k+1} \als^{k+1}}$ is a deviation in the prefactor from the central value $(\mu^2 / M^2)^{h_1 \als + h_2 \als^2 + \cdots + h_{k+1} \als^{k+1}}$, which is an $\co(\als^k)$ quantity in the EFT power counting.
This ``fundamental ambiguity in the exponentiation of the rapidity logarithms'' was first pointed out in \Ref{Chiu:2012ir}.
This ambiguity itself should be regarded as part of rapidity scale uncertainty,
and our form~\eq{factorization.expanded.asymm} (or~\eq{factorization.expanded}) 
 captures this ambiguity.

\section{An application: the $WW$ production with jet-veto}
\lb{s.application} 
Let us now apply the factorization formula~\eq{factorization.expanded} to the $WW$ production with jet-veto at $k=1$, i.e., at NNLL\@.
For the exponent coefficients $h_1$ and $h_2$, we need $\co(\als^2)$ calculations.
Combining the parametrization~\eq{parametrization} and solutions~\eq{g1g2} gives
\beq
h_1 
&=
\fr{1}{4\pi} ( d_1 + \Ga_0 \Lp)
\,,\\
h_2
&=
\fr{1}{(4\pi)^2} \!\lt[ d_2 + (\Ga_1 + \be_0 d_1) \Lp + \fr{\Ga_0 \be_0}{2} \Lp^2 \rt]
.\lb{e.g}
\eeq
For the $WW$ production at the 7- and 8-TeV LHC runs, 
the $q\bar{q}$ initial states dominate,
so the coefficients $d_{1,2}$ and $\Ga_{0,1}$ must be found for the fundamental representation of $\SU(3)$.
From explicit diagrammatic calculations, 
one finds $d_1 = 0$, while $d_2$ can be found in \Ref{Becher:2013xia}, where it is denoted by $d_2^{\text{veto}}$, while $\Ga_{0,1}$ and $\beta_0$ can be found in \Ref{Becher:2010tm}.  
The dependence of jet-veto on the jet-radius parameter $R$ enters the calculation through $d_2^{\text{veto}}$, and this $R$ dependence was first studied in \Refs{Banfi:2012yh, Tackmann:2012bt, Banfi:2012jm}~\cite{Becher:2013xia}.

For $\bigl[ \hat{\soft} \hat{B} \hat{\Bbar} (\mu) \bigr]$, 
we need $\co(\als)$ calculations.
From an explicit 1-loop calculation, one finds that $\hat{B}$ is given by
\beq
\hat{B} 
=
\sum_{i=q, g} \int_\xi^1 \fr{\dd{z}}{z} 
\hat{I}_{q \leftarrow i}(z, \ptv, \mu) \, \phi_i (\xi/z, \mu)
\,,  
\eeq
where $\phi_i$ is the PDF of parton $i$ for the proton and 
\beq
&\hat{I}_{j \leftarrow i}(z, \ptv, \mu)
\\
&=
\!\lt[ 1 -\as \!\lt( \Ga_0 \fr{\Lp^2}{4} + \ga_0 \Lp \rt)\rt] \! \de(1-z) \, \de_{ji}
\\
&\PE
-\as \!\lt[ \mathcal{P}_{j \leftarrow i}^{(1)} (z) \, \fr{\Lp}{2} - \mathcal{R}_{j \leftarrow i} (z) \rt],
\eeq
where $\mathcal{P}_{j \leftarrow i}^{(1)}$ and $\mathcal{R}_{j \leftarrow i}$ can be found
in Appendix A of \Ref{Becher:2014aya}.
$\hat{\Bbar}$ is given by the same formula except for the obvious replacement of ``$q$'' by ``$\bar{q}^{\,}$''.
Then, the whole $\hat{\soft} \hat{B} \hat{\Bbar}$ can be determined by matching it onto the corresponding full QCD expression at $\mu = \ptv$ 
and evolving it to another $\mu$ by using the $\mu$-RGEs~\eq{mu_RG} and~\eq{mu_RG.soft}. 
Alternatively, $\hat{\soft}$ can be directly calculated diagrammatically 
by regarding it as the soft function.
Either way, one finds $\hat{\soft} = 1$.

We can now update the scale uncertainty estimates for the $WW$ jet-veto cross-sections presented in \Ref{Jaiswal:2014yba}.
Note that the central values of the predictions, defined by $r=1$ and $\mu=\ptv$, 
are clearly unaffected.
However, scale uncertainties will necessarily be modified. In particular, 
the scale uncertainties arising from varying $r$ from $1/2$ to $2$ will be added in quadrature to those from varying $\mu$ from $\ptv/2$ to $2\ptv$, 
treating the two sources of uncertainties as independent.
We will not include power corrections as they were shown to be very small in \Ref{Jaiswal:2014cna} for the $WW$ production at the 7- or 8-TeV LHC run with 
$\ptv = 25$--$30$ GeV\@.
For other processes and/or choices of parameters, 
power corrections can become important, 
and a formalism for smoothly transitioning from the resummation-dominated regime to the fixed-order regime is described in \Ref{Stewart:2013faa}.
Including the uncertainties from non-perturbative effects~\cite{Becher:2013iya} \cite{Becher:2014aya} is beyond the scope of this paper.

In \Fig{R1.0_pi2}, the NLL and NNLL $WW$ jet-veto cross-sections are shown for the $\sqrt{s} = 8$ TeV LHC run with the anti-$k_\mathrm{T}$ jet algorithm~\cite{Cacciari:2008gp} with the jet-radius parameter $R = 1$. 
For the NLL, $h_2$ was discarded and $\hat{B}$ and $\hat{\Bbar}$ are set to the corresponding PDFs.
The `$\pi^2$ resummation'~\cite{Parisi:1979xd, Sterman:1986aj, Magnea:1990zb} 
is included in all the cases as in our earlier work~\cite{Jaiswal:2014yba}.%
\footnote{For a thorough discussion of scale uncertainty associated with `$\pi^2$ vs no $\pi^2$', see \Ref{Jaiswal:2014cna}.}
The plot on the right is based on the naive factorization formula as used in \Ref{Jaiswal:2014yba}, 
where we see that the NNLL uncertainty is ridiculously small ($\lsim 1\%$),
much smaller than the well-established NLO scale uncertainty of the \emph{inclusive} $WW$ cross-section, $\sim 3$--$4$\%~\cite{Campbell:2011bn}.
On the other hand, 
the plot on the left is obtained from the correct factorization formula~\eq{factorization.expanded},
which now exhibits scale uncertainties of reasonable size.   

In \Fig{R0.4_pi2}, we have repeated the computations for $R=0.4$, 
a typical value chosen in LHC experiments. 
Clearly, the plot on the left, which uses the correct factorization formula~\eq{factorization.expanded}, 
displays a better convergence of the NLL and NNLL predictions 
than the plot on the right, which is the one presented in \Ref{Jaiswal:2014yba}.

Comparing the two plots on the left, 
we see that 
the NNLL band in the $R=0.4$ case is much broader than that in the $R=1$ case.
This broadening is mainly due to a larger value of $h_2$ 
in \Eq{factorization.expanded}.
The large value of $h_2$ for $R=0.4$ comes from a large $d_2$ due to a $\log R$ multiplied by a large coefficient enhanced 
by factors of $C_\text{A}$ and $T_\text{F} n_\text{f}$ (further multiplied by numerical factors of $\co(10)$) arising from gluon splitting.%
\footnote{For a detailed study of $R$ dependence at $\co(\als^3)$ and the resummation of $\log R$, see \Refs{Alioli:2013hba}~and~\cite{Dasgupta:2014yra}, respectively.
Including the results of those studies into our analysis is beyond the scope of the present paper.}
Perhaps more importantly, note that a modest change in $h_2$ can lead to substantial 
change in the final numerical results as $h_2$ appears in the exponent of the ratio $\mu/M$.
Not having $h_2$, on the other hand, 
the NLL bands of the $R=1$ and $R=0.4$ cases have a similar width.

\begin{figure}[t]
\includegraphics[width=0.49\linewidth]{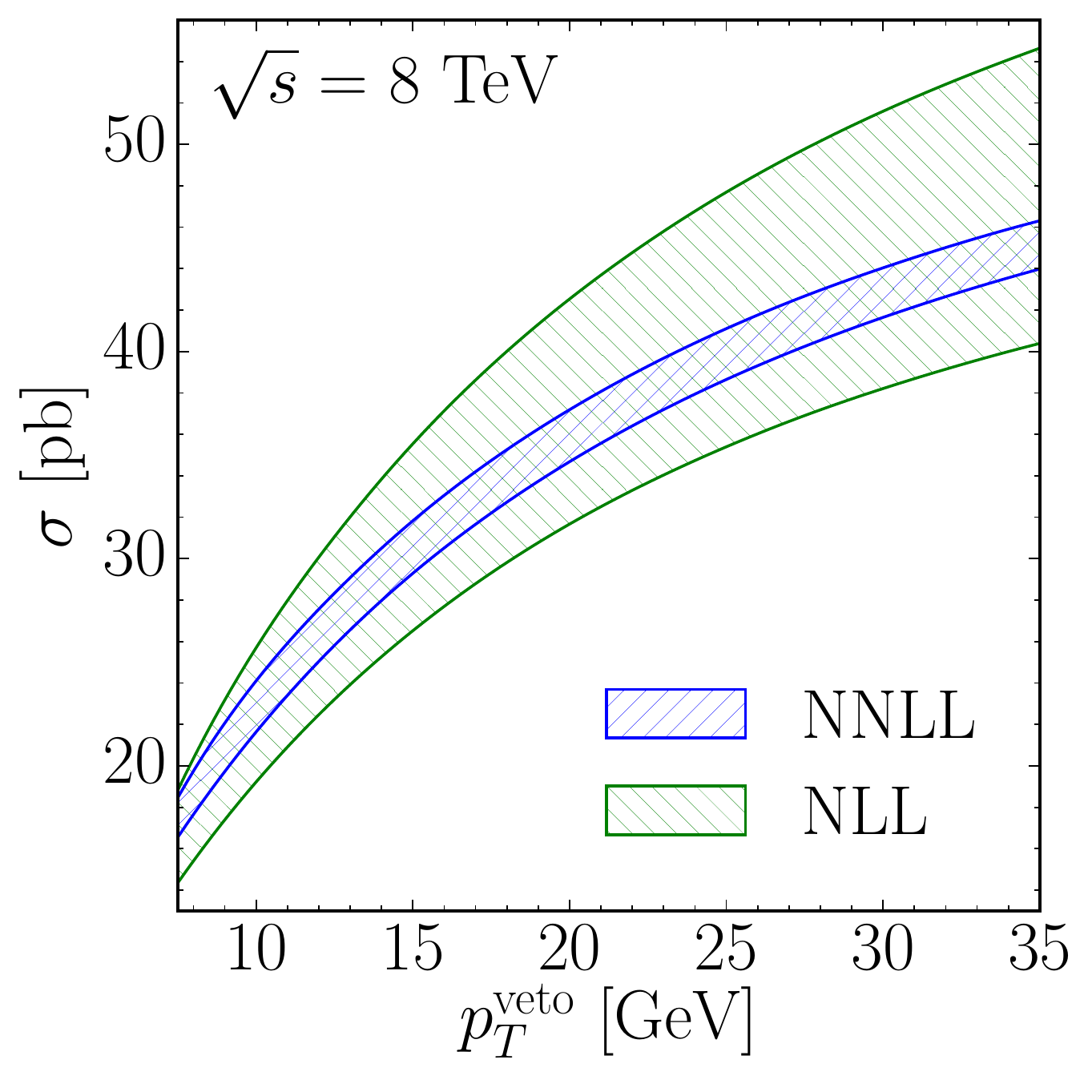} 
\includegraphics[width=0.49\linewidth]{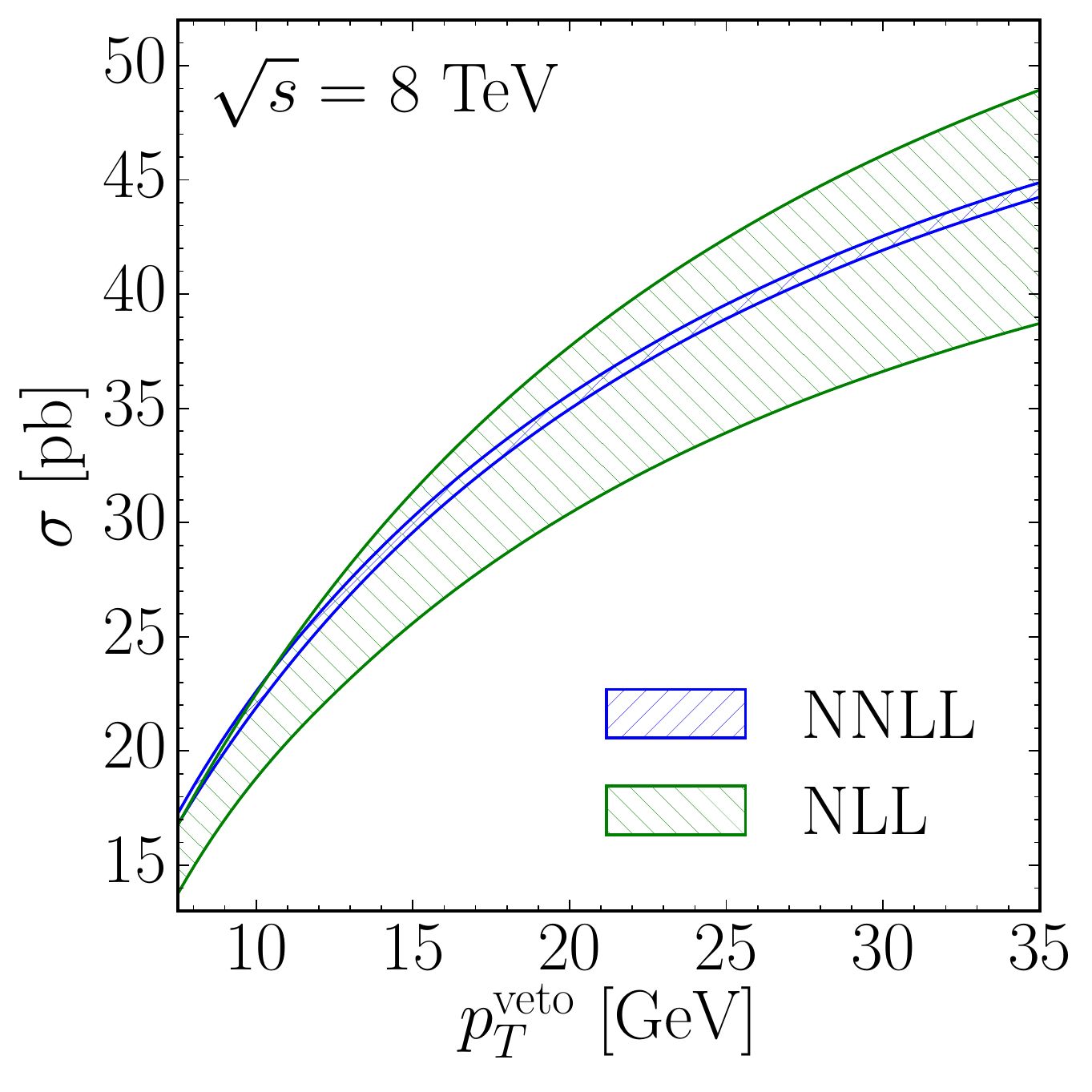}
\caption{The NLL and NNLL jet-veto cross-sections for $WW$ production for the 8-TeV
LHC run with $R=1$. The scale-uncertainty bands in the left plot are obtained from 
the correct factorization formula~\eq{factorization.expanded}, 
while in the right-hand plot, they are obtained following~\cite{Jaiswal:2014yba}. } 
\lb{f.R1.0_pi2}
\end{figure}
\begin{figure}[t]
\includegraphics[width=0.49\linewidth]{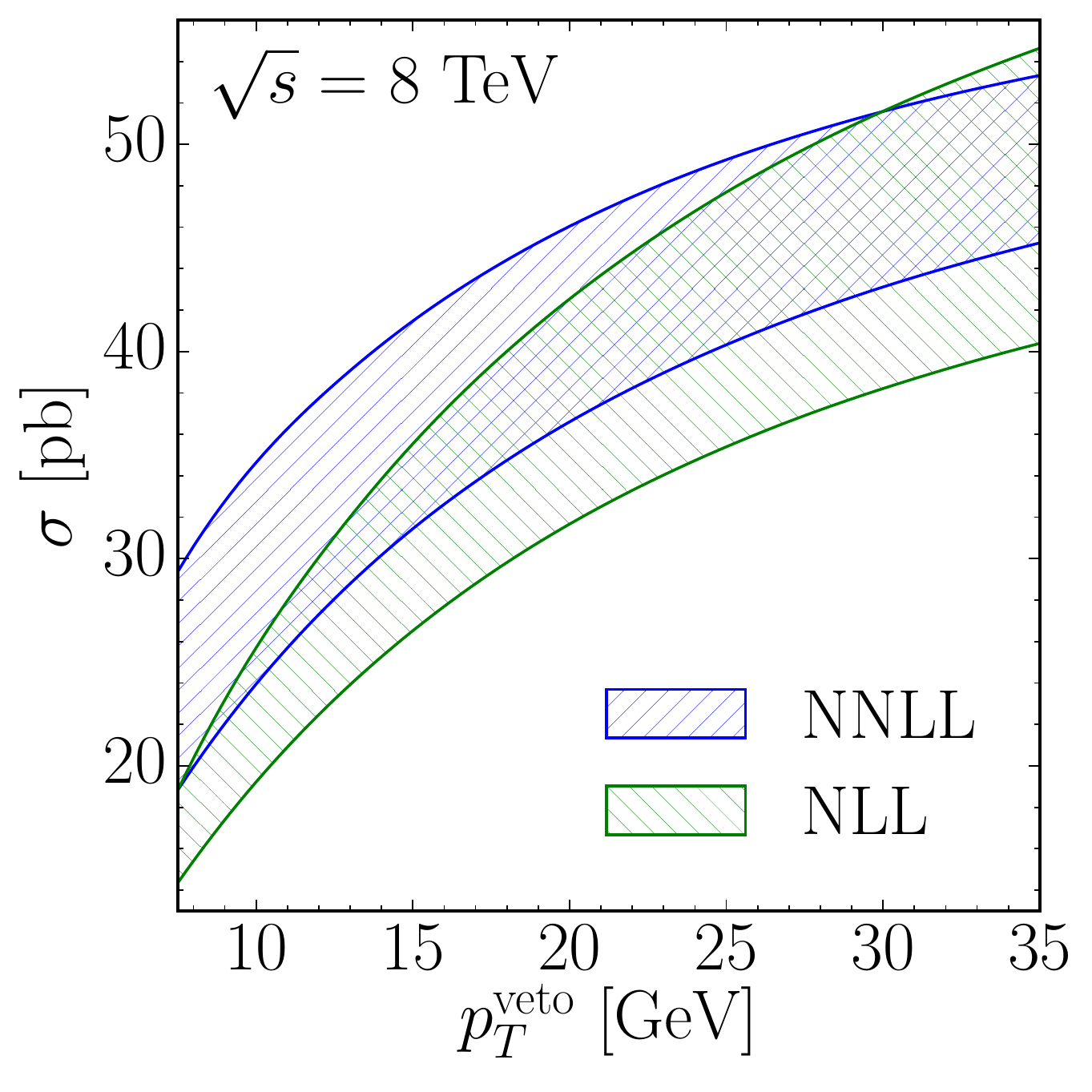} 
\includegraphics[width=0.49\linewidth]{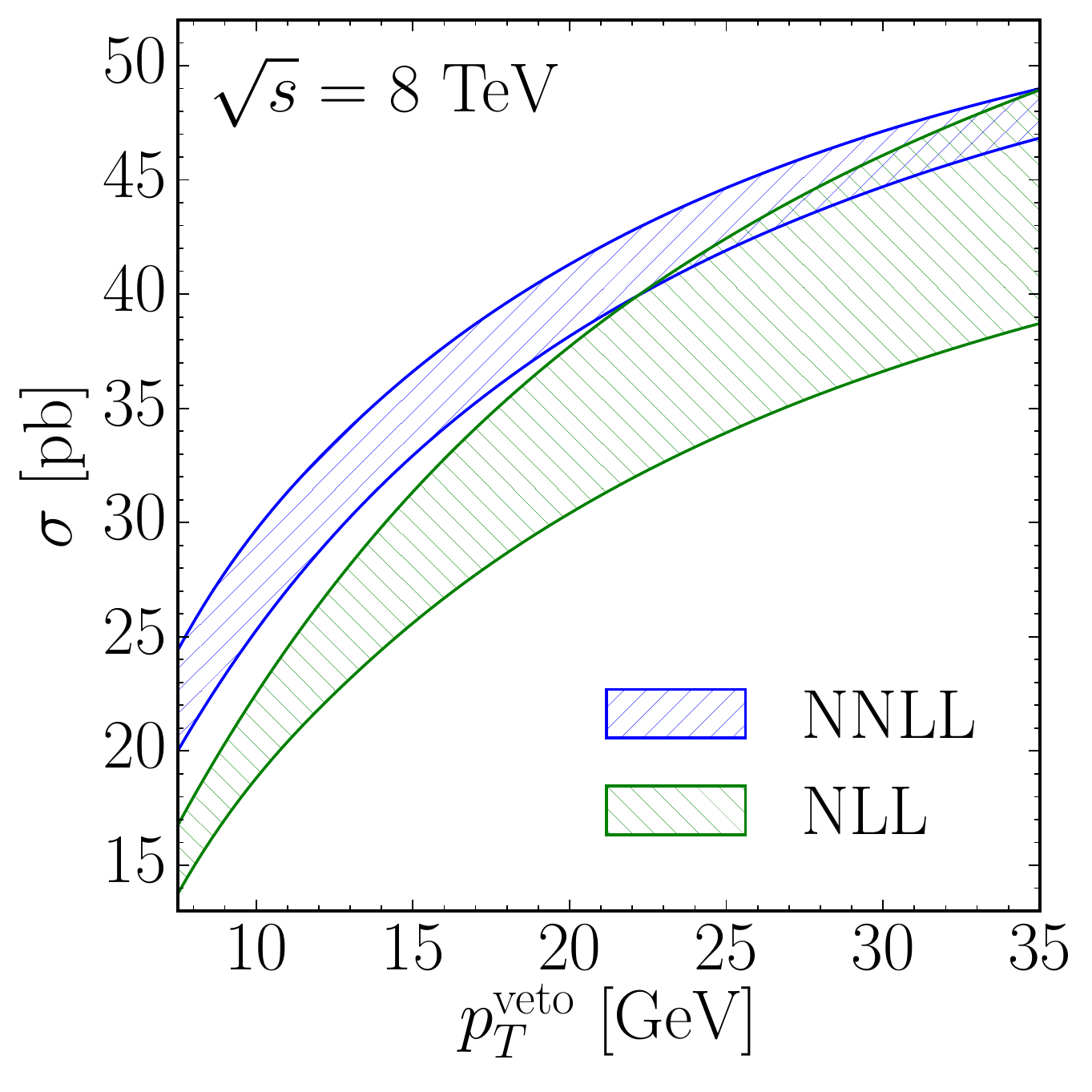}
\caption{Same as \Fig{R1.0_pi2} but for $R=0.4$.}
\lb{f.R0.4_pi2}
\end{figure}
%

\begin{acknowledgments}
PJ thanks Patrick Meade for discussions on scale uncertainties in the $WW$ jet-veto cross-sections.
We also thank our anonymous referee for pointing out to us that our previous version of \Eq{factorization.expanded} was only sensitive to the single coefficient $h_{k+1}$ and thus overly simplistic as discussed in \Sec{truncation}\@.
TO is supported by the US Department of Energy under grant DE-FG02-13ER41942.

\end{acknowledgments}

\appendix

\section{An analytic regulator with a collinear-anti-collinear asymmetry}
\lb{a.asymmetric}
Instead of the analytic regulator~\eq{ana_reg}, 
we can also use a different analytic regulator where we use
\beq
 \lt( \fr{\nu}{k_+}    \rt)^{\!\! \al}
\lb{e.asymm.reg}
\eeq 
for \emph{both} the collinear and anti-collinear sectors~\cite{Becher:2011dz, Becher:2012qa},
which can be quite convenient for practical calculations due to its simplicity.
While the asymmetric treatment of the collinear and anti-collinear sectors means 
we cannot obtain $\Bbar$ from $B$ by 
simply relabelling as $\ptv \to \ptvbar$, etc., 
this asymmetric regulator has the simplicity that it only introduces one rapidity scale, $\nu$.
Since the action of the asymmetric regulator~\eq{asymm.reg} in the collinear sector is the same as the symmetric case~\eq{ana_reg}, 
the form of $L$ is unmodified from the form~\eq{L_and_Lbar}.
On the other hand, in the anti-collinear sector, the presence of $(\nu / k_+)^\al$ instead of $(\nubar / k_-)^\albar$ implies that $\Lbar$ should be changed to
\beq
\Lbar' \equiv \log \fr{\nu \xibar P_-}{\mu^2}
\,.\lb{e.asymm.Lbar}
\eeq
This form can be understood from the fact that $k_+ = k_\mathrm{T}^2 / k_-$ (with $k_\mathrm{T} \equiv |\vec{k_\perp}|$), 
where the $k_\mathrm{T}$ integration is cut off by dimensional regularization at the scale $\mu$
while the  $k_-$ integration picks up $\xibar P_-$, the only physical scale in the anti-collinear sector.         

Then, repeating a similar analysis as we did with the symmetric regulator in \Sec{derivation}, 
we find that the $\mu$-RGEs~\eq{mu_RG} and~\eq{mu_RG.soft} are modified as 
\beq
\mu \fr{\del}{\del\mu} \log B     (\mu, \nu)
&=  
f^{(1)\!}(\als) \,  L      + f^{(0)\!} (\als, \Lp) 
\,,\\
\mu \fr{\del}{\del\mu} \log \Bbar (\mu, \nu)
&=  
-f^{(1)\!}(\als) \, \Lbar' + \bar{f}^{(0)\!} (\als, \Lpbar) 
\,,\\
\mu \fr{\del}{\del\mu} \log \soft (\mu, \nu)
&= \wtd{F}^{(0)\!}(\als, \Lp, \Lpbar)
\,,\lb{e.mu_RG.asymm}
\eeq
where the $B$ equation is the same as before as nothing has changed in the collinear sector, while the $\Bbar$ and $\soft$ equations are changed. Due to the lack of symmetry, the function $\fbar^{(0)}$ is not a priori the same as $f^{(0)}$,
and $\wtd{F}_0$ has no reason to be the same as $F^{(0)}$ either. 
Also observe that the right-hand side of the $\soft$ equation above has no dependence on $\nu$. 
The constraints~\eq{f1} and~\eq{f0} remain as before
with the obvious replacements $f^{(0)}(\als, \Lpbar) \to \bar{f}_0(\als, \Lpbar)$ and 
$F^{(0)} \to \widetilde{F}^{(0)}$.

For rapidity RGEs, we find that the $\nu$-RGEs~\eq{nu_RG.beam} and~\eq{nu_RG.soft} are modified as
\beq
\nu    \fr{\del}{\del\nu} \log B     (\mu, \nu)
&=  
\sum_{p=0}^\infty g^{(p)\!} (\als) \, \Lp^p  
\,,\\
\nubar \fr{\del}{\del\nu} \log \Bbar (\mu, \nu) 
&=  
-\sum_{p=0}^\infty g^{(p)\!} (\als) \, \Lpbar^p
\,,\\
\nu    \fr{\del}{\del\nu} \log \soft (\mu, \nu)
&=  
-\sum_{p=0}^\infty g^{(p)\!} (\als) \, (\Lp^p - \Lpbar^p) 
\,.\lb{e.nu_RG.asymm}
\eeq
Again, the $B$ equation is unmodified.
The negative sign in the $\Bbar$ equation here is a consequence of 
the negative sign in the $\Bbar$ equation in~\eq{mu_RG.asymm}
combined with a relation similar to \Eq{path_independence} from path-independence requirement in the $\mu$-$\nu$ space.

Solving \Eq{nu_RG.asymm} 
for $B$ from $\nuinitial$ to an arbitrary final $\nu$,
$\Bbar$ from $\nubarinitial$ to $\nu$, 
and $\soft$ from $\nusinitial$ to $\nu$
we find that the factorization formula~\eq{factorization.original} is modified as
\beq
& \soft(\mu, \nu) \, B(\mu, \nu) \, \Bbar(\mu, \nu)
\\
&= 
\!\lt( \fr{\nusinitial}{\nuinitial} \rt)^{\! g(\mu)} \!
\!\lt( \fr{\nubarinitial}{\nusinitial} \rt)^{\! \gbar(\mu)} \!
\hat{\soft} \hat{B} \hat{\Bbar} (\mu)
\lb{e.factorization.original.asymm}
\eeq
with $\hat{\soft} \hat{B} \hat{\Bbar} (\mu) 
\equiv 
\soft(\mu, \nusinitial) \, 
B(\mu, \nuinitial) \, \Bbar(\mu, \nubarinitial)$.
We can again show `$r$ invariance' as we did in \Sec{truncation},
and the minimization of $\Lpbar'$ tells us that the choice for $\nubarinitial$ in \Eq{nu_choice} should be modified to
\beq
\nubarinitial = \fr{\mu^2}{\xibar P_-}
\,,
\eeq
while $\nuinitial$ remains unchanged.
For $\mu$ and $\nusinitial$, we again see $\nusinitial \sim \mu \sim \ptv$ 
by matching $\hat{\soft} \hat{B} \hat{\Bbar}$ to the corresponding full QCD quantity,
since $\hat{\soft} \hat{B} \hat{\Bbar}$ (by definition) only depends on $\mu$, $\nusinitial$, and $\ptv$, 
while the QCD counterpart has only $\mu$ and $\ptv$.
Alternatively, if we integrate in the soft modes in the SCET and regard $\soft$ as the soft function, it is clear that $\ptv$ is the only physical scale in the problem and hence $\mu \sim \nusinitial \sim \ptv$.
Hence, we choose $\nusinitial = \mu$ as in the symmetric regulator counterpart~\eq{nu_choice}.

For the case of practical interest $\ptvbar = \ptv$, 
we have $\gbar = g$, which makes $\nusinitial$ cancel out in \Eq{factorization.original.asymm},
and the ratio $\nubarinitial / \nuinitial$ becomes $\mu^2 / M^2$  
as in the symmetric regulator case.
Furthermore, from the $\nu$-RGE for $\soft$ in \Eq{nu_RG.asymm},
we see that $\soft$ becomes $\nu$ independent when $\gbar = g$ 
(while it may still depend on $\mu$).
Therefore, the factorization formula~\eq{factorization.original.asymm} reduces
to \Eq{factorization.asymm},
which is essentially identical to the symmetric regulator 
counterpart~\eq{factorization.exact}.
The discussion of rapidity scale uncertainty in \Sec{results} 
then goes through essentially unmodified and leads to
the final factorization formula~\eq{factorization.expanded.asymm}.

%

\end{document}